\documentclass{aa}
\usepackage[fleqn]{amsmath}
\usepackage{graphicx}
\usepackage{natbib}
\bibpunct{(}{)}{;}{}{} 

\DeclareMathOperator{\curl}{{rot}}
\DeclareMathOperator{\Div}{div}
\DeclareMathOperator{\grad}{grad}
\DeclareMathOperator{\vnabla}{\vec\nabla}
\newcommand\Msun{M_{\odot}}

\begin{document}

\title{Resonance accretion on the magnetized\\ rotating neutron star}
\titlerunning{Resonance accretion}

\author{Ya. N. Istomin\inst1 \and A. V. Makarov\inst2}

\offprints{Ya. N. Istomin}

      \institute{
         P.~N.~Lebedev Physical Institute,
         Leninsky Prospect 53, Moscow, 117924 Russia\\
         \email{istomin@lpi.ru}
      \and
         Department of General and Applied Physics,
         Moscow Institute of Physics and Technology,\\
         Institutsky 9, Dolgoprudny, Moscow region, 141700 Russia\\
         \email{stalker@dgap.mipt.ru}
      }

\date{Received \dots/ Accepted \dots}

\abstract{
It is suggested that the accretion disk excites
the eigen modes of Alfven oscillations of the magnetic field
tubes the ends of which are frozen to the neutron star surface.
The resonance takes place when the eigen Alfven frequency
coincides with the difference of the Keplerian disk rotation
frequency and the frequency of rotation of a neutron star.
This model explains the two narrow lines of QPO oscillations in
kHz band and permits to calculate the masses and radii of
the neutron stars in X-ray binaries.
   \keywords{Accretion, accretion disks -- Waves -- Stars: neutron --
   X-rays: binaries -- Stars: individual: Sco X-1}
}

\maketitle

\section{Introduction}

Rossi X-Ray Timing Explorer (RXTE) in 1996 discovered the quasi-periodic
oscillations (QPO) of X-ray emissions in low-mass X-ray binaries (LMXB)
in the band of $1$~kHz frequencies \citep{klis2,klis1}.
They are narrow, the
ratio of the line's width to the central frequency can be as low as
$10^{-2}$, and in most cases there were observed two lines.
The existence of such
lines means that the accretion disk approaches very close to the surface of
the rotating neutron star. Indeed, the Keplerian frequency
\begin{equation}
\nu_{\mathrm{K}}=\frac{1}{2\pi }\left(\frac{MG}{r^{3}}\right)^{1/2},
\end{equation}
where $M$ is the neutron star's mass and $r$ is the distance from
the star's center, is of the order of 1~kHz at the distance of $1.7\cdot
10^{6}$~cm for the typical neutron star mass $M\simeq 1.4\cdot
\Msun$. But estimated neutron star radius $R$ is of about $10^{6}$~cm,
and we see that the inner edge of the accretion disk can approach almost to
the star surface. If we estimate the kinetic energy of a proton near the
star's radius ($r\simeq R$) it turns out of $3\cdot 10^{-5}$~erg or 20~MeV.
For the characteristic proton's density in the accretion flow near the star
surface, corresponding to the accretion rate $\dot M\simeq 10^{17}$~g/sec
($L_x\simeq L_{\mathrm{Edd}}$),
$n_{\mathrm p}\simeq 8\cdot 10^{17}$~cm$^{-3}$, we obtain the
energy density of the accreting plasma of
\begin{equation*}
{\mathcal E}\simeq 2.5\cdot 10^{13}~\mathrm{erg/cm}^{3}.
\end{equation*}
It seams that for the existence of the disk in the region near the star's
surface, the star's magnetic field should not exceed the value of
\begin{equation*}
\frac{B^{2}}{8\pi }<{\cal E},\quad \mathrm{or}\quad B<2.5\cdot 10^{7}~\mathrm{G}.
\end{equation*}
But this low value of magnetic field on the neutron star surface contradicts
with the magnitude of the magnetic field observed for the milliseconds pulsars
$B\simeq 10^{9-10}$~G, which are considered originate from the LMXB. But these
estimations are incorrect. The plasma destroys the magnetic field structure
when its thermal energy exceeds the magnetic field energy, but not when its
kinetic flow energy exceeds the magnetic one. Moving across the magnetic
field lines plasma polarizes so that the electric field across the magnetic
one appears. In the case of a thin disk the radial electric field
\begin{equation}
{\vec E_{r}}=-\frac{1}{c}\lbrack {\vec v_{\mathrm{K}}}{\vec B}\rbrack
\label{eq2}
\end{equation}
is enough to carry the flow across the magnetic field with the Keplerian
velocity $v_{\mathrm{K}}=2\pi \nu _{\mathrm{K}}r$. For the dipole magnetic field ($B\propto
r^{-3}$) the disk of the temperature $T$ destroys the star's magnetic field
at the distances
\begin{equation}
r>r_{A}=R\left[ \frac{B^{2}R^{2}}{4\dot{M}(T/m_{\mathrm p})^{1/2}}\right] ^{1/4}
\label{eq3}
\end{equation}
If we substitute to \eqref{eq3}, the characteristic energy of X-ray emission
($\sim1$~keV) as the maximum disk temperature we obtain
\begin{equation*}
r>20R\left(\frac{B}{10^{9}~\mathrm{G}}\right)^{1/2}.
\end{equation*}
Thus, we see that in the region where QPO of 1 kHz frequency can originate
the structure of the magnetic field of the star is not disturbed by the
accreting plasma. We can consider that the structure of the magnetic field near
the star surface($r\leq 10R$) is given by the interior of the neutron
star and has the dipole configuration. Besides, the accretion changes the
initial neutron star magnetic field $B\simeq 10^{12}$~G
up to $B\simeq 10^{9}$~G.
And it is naturally to consider the direction of the dipole magnetic
momentum $\vec\mu$ coinciding with the axis of star's rotation
$\vec\Omega$.

\begin{figure}
\includegraphics[width=88mm]{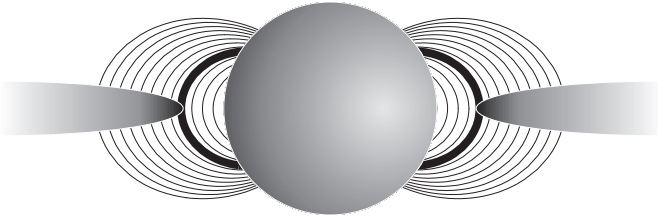}
\caption{The accreting disk approaches so close to
the neutron star surface, that plasma flow along the magnetic field lines.}
\label{accrstar}
\end{figure}

We assume that the accreting disk approaches close to the neutron
star surface.
Plasma flows from the disk to the neutron star along magnetic field lines
and form so-called ``magnetic field plasma tubes'', which are shaped as
magnetic field lines, are filled with plasma and have a small thickness.

The narrow structure of the QPO of 1 kHz frequency suggests the resonance
nature of its origin. We have the idea that it is a resonance between
Keplerian rotation of the disk and an eigen modes of the oscillation of
magnetic field tubes, frozen to the conducting star surface.

Such model is more close to the sonic  point beat frequency
model \citep{miller}, where the resonance takes place not with
the Keplerian rotation of the disk, but with the radial motion
of accreting flow coinciding with the sonic velocity.
The other model based on the eigen modes of the disk
in the rotating frame was proposed by \cite{titarchuk}.
There is more exotic mechanism proposed by \cite{klein}
where QPOs are the oscillations of photon's bubbles.

\section{Magnetic tubes oscillations}

In this section we discuss the
oscillations of the magnetic field lines with  plasma frozen to them.
Discovering this model it is convenient to
use the dipole coordinates ($l,a$) introduced by the definition:
\begin{eqnarray}
{\vec B} &=&-\vnabla l;  \nonumber \\
{\vec B} &=&[\vnabla a,\vnabla \phi].
\end{eqnarray}
$l$ is the coordinate along the magnetic field line
\begin{equation}
l=\frac{({\vec{\mu}}{\vec r})}{r^{3}}=\mu \frac{z}{(\varrho ^{2}+z^{2})^{3/2}},
\end{equation}
$a$ numerates the magnetic field surfaces
\begin{equation}
a=\mu \frac{\varrho ^{2}}{(\varrho ^{2}+z^{2})^{3/2}}.
\end{equation}
$a=\mathrm{const}$ is the equation for the magnetic field lines
\begin{equation}
z^{2}=(\frac{\mu }{a})^{2/3}\varrho ^{4/3}-\varrho ^{2},\quad \text{or}\quad r=%
\frac{\mu }{a}\sin ^{2}\theta .
\end{equation}
Here $(\varrho, z)$ are the cylindrical coordinates, $(r,\theta)$ are the
spherical ones. The value of $a$ is also connected with the maximum deflection
$\varrho_{max}$ of the dipole magnetic field line from the star's center
$ a=\mu /\varrho _{max}$. For different magnetic surfaces the
coordinate $a$ changes from 0 to $\mu /R$. The coordinate $l$ changes for a
given $a$, from $-(\mu ^{2}-\mu R a)^{1/2}/R^{2}$ to $(\mu
^{2}-\mu R a)^{1/2}/R^{2}$. The point $l=0$ corresponds to the top of the magnetic
field line. Three coordinates $(l,a,\phi)$ ($\phi $ is the azimuthal
angle) are the curvilinear orthogonal coordinate system.

Because in the near ($r<10R$) region $B^{2}/8\pi \gg{\cal E}$
the plasma can be considered as cold, the plasma density is
proportional to the magnetic field strength, $n_{\mathrm p}=\lambda B$. Here
$\lambda$ is the constant not depending on the coordinates. The equations of
the ideal MHD in rotating with angular velocity $\Omega$ frame ($\Omega$
is the star rotation frequency) can be presented for the electric current
${\vec j}$ as follows
\begin{equation}
{\vec j}=\frac{v_{A}^{2}}{\omega ^{2}B}\curl\curl\left\{ \frac1B
\Bigl[\vec B[\vec j\vec B] \Bigr] \right\},
\end{equation}
where $v_{A}$ is the Alfven velocity
\begin{equation*}
v_{A}^{2}=B^{2}/4\pi n_{\mathrm p}m_{\mathrm p}=B/4\pi \lambda m_{\mathrm p}.
\end{equation*}
We consider here that all the perturbed values $\vec j$,
$\delta\vec B$, $\vec E$
depend on the time $t$  and the azimuthal angle $\phi $  as $\exp
\left\{ -i\omega t+im\phi \right\} .$

The electric current can be presented through its covariant components
$j_{1},\,j_{2},\,j_{3}$
\begin{equation*}
{\vec j}=j_{1}\vnabla l+j_{2}\vnabla a+j_{3}\vnabla \phi ,
\end{equation*}
for which from (8) we obtain
\begin{gather}
\begin{split}
j_{2}\left(1-\frac{v_{A}^{2}}{\omega ^{2}\varrho ^{2}}m^{2}\right) +\frac{%
v_{A}^{2}}{\omega ^{2}\varrho ^{2}B}\frac{\partial }{\partial l}\left[
\varrho ^{2}B^{2}\frac{\partial }{\partial l}\left(j_{2}B\right) \right] =\\
im\frac{v_{A}^{2}}{\omega ^{2}\varrho ^{2}B}\frac{\partial }{\partial a}\left(
j_{3}B\right);
\label{eq9}
\end{split}
\\
\begin{split}
j_{3}+\frac{v_{A}^{2}\varrho ^{2}B}{\omega ^{2}}\left[ \frac{\partial ^{2}}{%
\partial a^{2}}\left(j_{3}B\right) +\frac{\partial }{\partial l}\frac{1}{%
\varrho ^{2}}\frac{\partial }{\partial l}\left(j_{3}B\right) \right] =\\
im\frac{v_{A}^{2}\varrho ^{2}B}{\omega ^{2}}\frac{\partial }{\partial a}\left(
j_{2}B\right) ;
\label{eq10}
\end{split}
\\
j_{1}=\frac{v_{A}^{2}}{\omega ^{2}B}\left[ \frac{\partial }{\partial a}%
\varrho ^{2}B^{2}\frac{\partial }{\partial l}\left(j_{2}B\right) +\frac{im}{%
\varrho ^{2}}\frac{\partial }{\partial l}\left(j_{3}B\right) \right].
\label{eq11}
\end{gather}

The boundary conditions for (\ref{eq9}, \ref{eq10}, \ref{eq11}) are on the star surface $r=R$ . Due to the
ideal conductivity of neutron star matter the tangential components of the
electric field have to be zero. The electric field of the disturbance is
proportional to the electric current
\begin{equation*}
{\vec E}=\frac{iB}{\omega c^{2}\lambda }\left(j_{2}\vnabla a+j_{3}\vnabla \phi
\right) .
\end{equation*}
It gives the boundary conditions for $j_{2}$ and $j_{3}$
\begin{multline}
j_{2}=j_{3}=0\text{ at } r=R,\text{ or }\\
l=\pm l_m =\pm\left(\mu^2-\mu R a\right) ^{1/2}/R^{2}.
\end{multline}

Equation \eqref{eq9} describes the Alfven type oscillations because it contains only
the second derivatives over longitudinal coordinate $l$ , but equation
\eqref{eq10}
is that for the magnetosonic waves (also the derivatives over $a$
coordinate). In the case of nonaxial symmetric oscillations ($m\neq 0$)
these modes are coupled. We consider the case when the accretion disk
excites the MHD waves in the magnetosphere with the dipole magnetic field.
Rotating with the Keplerian frequency $\omega _{\mathrm{K}}$ different from the star
rotation frequency $\Omega $, the conducting disk deflects the magnetic field
lines in equatorial plane to $\phi $-direction. Because this deflection is
symmetric over $\phi $ we consider here only the case $m=0$. As
Alfven and magnetosonic oscillations are independent and Alfven wave has
only the $\phi $ component of the disturbed magnetic field
\begin{gather*}
j_{2}=-\frac{c}{4\pi \varrho ^{2}}\frac{\partial \delta B_{3}}{\partial l},\\
\delta {\vec B}=\delta B_{1}\vnabla l+\delta B_{2}\vnabla a+\delta B_{3}\vnabla
\phi ,\quad \delta B_{3}=\varrho \cdot \delta B_{\phi },
\end{gather*}
while the magnetosonic has two, $\delta B_{1}$ and $\delta B_{2}$
\begin{equation}
j_{3}=\frac{cB^{2}\varrho ^{2}}{4\pi }\left(\frac{\partial \delta B_{2}}{%
\partial l}-\frac{\partial \delta B_{1}}{\partial a}\right) .
\label{eq13}
\end{equation}
For this reason we choose only the Alfven wave ($\delta B_{1}=\delta
B_{2}=0,\,\delta B_{\phi }\neq 0$)
\begin{equation}
j_{2}+\frac{v_{A}^{2}}{\omega ^{2}\varrho ^{2}B}\frac{\partial }{\partial l}%
\left[ \varrho ^{2}B^{2}\frac{\partial }{\partial l}\left(j_{2}B\right)
\right] =0. \label{maineq}
\end{equation}

\begin{figure}
\includegraphics[width=88mm]{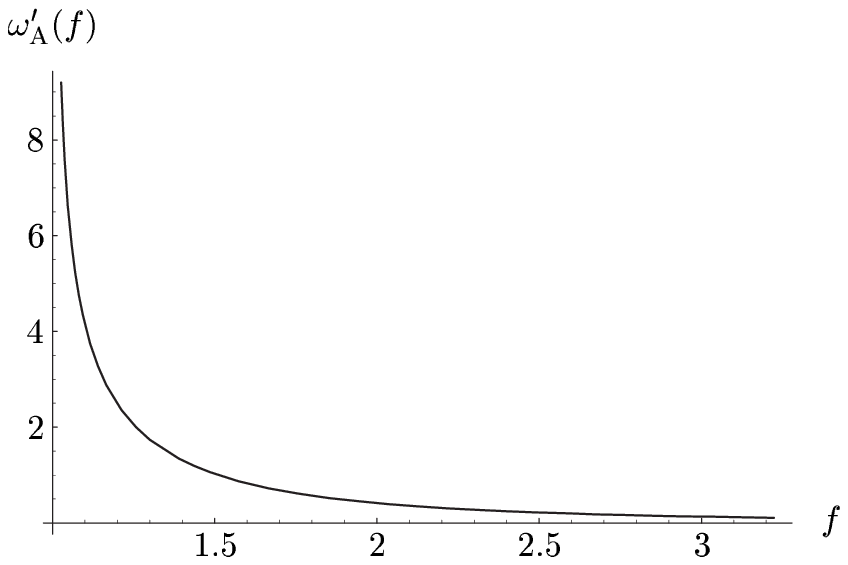}
\caption{The dependence of Alfven first harmonics $\omega'_{\mathrm A}$ on $f$.}
\label{alfven}
\end{figure}

The boundary conditions
\begin{equation*}
j_{2}\left(a,\,l=\pm l_m \right)
=0
\end{equation*}
imply the quantization of eigen frequencies of the Alfven modes
\begin{equation*}
\omega =\omega _{j}(a),\quad j=1,2,\ldots
\end{equation*}
After parameterization the eigen frequencies can be presented in the form
\begin{equation}
\omega _{j}=\omega _{0}\cdot \omega _{j}^{\prime}(f),
\end{equation}
where
\begin{equation*}
\omega _{0}=\left(\frac{\mu }{4\pi R^{5}\lambda m_{\mathrm p}}\right) ^{1/2},
\end{equation*}
is the characteristic Alfven frequency, and $\omega
_{j}^{^{\prime }}$ is the dimensionless amplitude depending on the
dimensionless parameter $f=\varrho _{\max }/R$, which means the height of
the dipole magnetic field line in the star radius units. The calculation
for the general mode $j=1$ gives the dependence of on $f$ as pictured on
Fig.~\ref{alfven}. At large $f$ the eigen frequency falls down like
$\omega_{\mathrm A}^{^{\prime }}\propto f^{-5/2}$.
The reason of this behavior is that the
main contribution to the frequency gives the wave structure in the top of
magnetic field tube. Indeed, multiplying the \eqref{maineq} on to
$\left(B^{2}\varrho^{2}\omega ^{2}/v_{A}^{2}\right) \cdot j_{2}^{\ast }$
and integrating it
over $l$ from$-l_{m}$ to $+l_{m}$, we get
\begin{equation}
\omega ^{2}=\frac{\int_{-l_{m}}^{l_{m}}\varrho ^{2}\left| B\frac{\partial }{%
\partial l}\left(Bj_{2}\right) \right| ^{2}\mathrm dl}{\int_{-l_{m}}^{l_{m}}\frac{%
\varrho ^{2}}{v_{A}^{2}}\left| \left(Bj_{2}\right) \right| ^{2}\mathrm dl}.
\label{eq16}
\end{equation}

From \eqref{eq16} it is seen that $\omega \simeq \left(v_{\mathrm A}\right) _{\min }/L$,
where $L$ is the length of the magnetic field line. Because of $v_{A}=\left(
B/4\pi \lambda m_{\mathrm p}\right) ^{1/2}\propto f^{-3/2}$ and $L\propto f$ we
obtain the calculated dependence
\begin{equation*}
\omega \propto f^{-5/2}\quad (f\gg1).
\end{equation*}
As for the values of $f$ near the star surface $|f-1|\ll 1$, then $L\propto
(f-1)^{1/2}$ and $\omega \propto (f-1)^{-1/2}\quad (f\simeq 1)$ as imagined
on the Fig.~\ref{alfven}. For the same reason the more high modes have the frequencies
very close to the harmonics of the general mode
\begin{equation}
\omega _{j}\simeq j\cdot \omega _{1}, \quad
\omega_{\mathrm A}\equiv\omega_1.
\end{equation}

For the plasma density of $n_{\mathrm p}\simeq 8\cdot 10^{17}$~cm$^{-3}$ near the
star surface and magnetic momentum $\mu \simeq 10^{27}$~G$\cdot$cm$^{3}$
($B\simeq 10^{9}$~G)
the frequency $\omega _{\mathrm A}$ is of the order of $2\pi$~kHz at the
distance $r\simeq 5.6R$. It means that QPO of 1 kHz can appear as the
result of the resonance of the disk rotation and Alfven eigen modes excited
by the disk
\begin{equation}
i\cdot (\omega _{\mathrm{K}}-\Omega)=
\omega _{j}\simeq j\cdot \omega _{\mathrm A}.
\label{resonance}
\end{equation}
Here $i$ is the integer number. The resonance condition \eqref{resonance}
defines the
magnetic field tubes, where the resonance occur, $f_{i,j}$ :
\begin{equation}
\omega _{\mathrm{K}}(f_{i,j})-\Omega =\frac ji\omega _{\mathrm A}(f_{i,j}).
\end{equation}

\begin{figure}
\includegraphics[width=88mm]{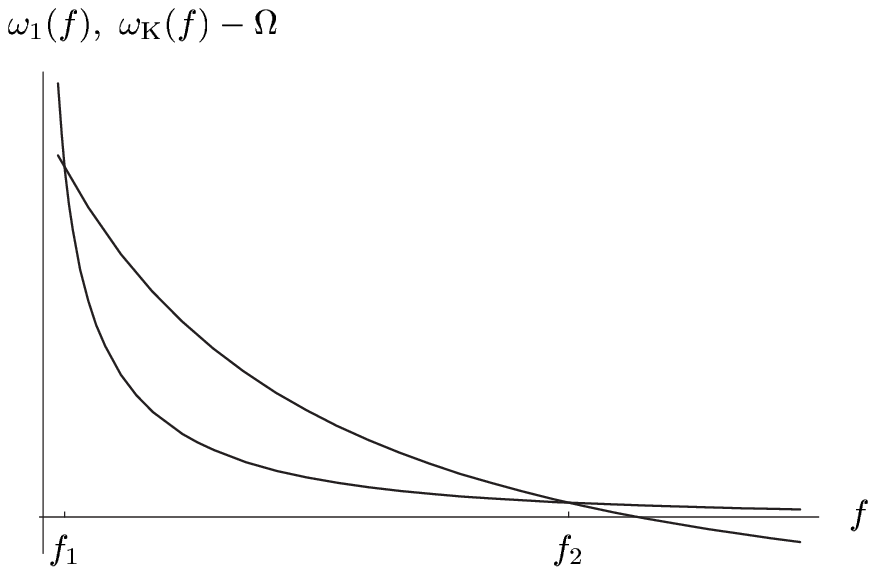}
\caption{There are two general solutions
$f_1$ and $f_2$ of the \eqref{resonance}.}
\label{alfkep}
\end{figure}

The solution of \eqref{resonance} is pictured on the Fig.~\ref{alfkep}.
In general case we have two solutions $f_{1}$ and $f_{2}$\quad ($f_{2}>f_{1}
$\thinspace). Their positions are not constants in time because of the
changing of the accretion rate and as a consequence of changing of the
plasma density $n_{\mathrm p}$ . The frequency of the Alfven eigen modes
$\omega_{j}\propto n_{\mathrm p}^{-1/2}$.
It means that higher accretion rate (or $n_{\mathrm p}$)
larger $f_{1}$ and smaller $f_{2}$. Accordingly $\omega _{\mathrm{K}}(f_{1})$
becomes larger, but $\omega _{\mathrm{K}}(f_{2})$ becomes smaller. The observed
changes of QPO frequency correspond to the behavior of the root $f_{1}$ .
Apart, it is observed the clear proportionality between intensity of X-ray
emission and the frequency of QPO of 1 kHz band. We consider that the
resonance at $f=f_{1}$ is more strong than at $f=f_{2}$ because of
\begin{itemize}
\item it is more close to the star surface where the dipole magnetic field is
large; and
\item usually $f_{2}$ is large enough, so that $\omega _{\mathrm{K}}(f_{2})$ falls down
from the 1~kHz frequencies.
\end{itemize}

Further we will discuss only the high frequency
solution $f=f_{1}$ . How the QPO frequencies are connected with discussed
resonance? At first the excited Alfven oscillations will destroy the
accretion disk. The magnetic field gives the drift velocity $\delta v_{z}$
for the disk plasma particles
\begin{equation}
\delta {\vec v}_{z}=c\frac{\left[ {\vec E}_{r}\,\delta {\vec B}_{\phi }\right]
}{B^{2}};\quad \mid \delta v_{z}\mid =v_{\mathrm{K}}\frac{\delta B_{\phi }}{B},
\end{equation}
(see ${\vec E}_{r}$ from \eqref{eq2}). The disk matter will be pushed from equatorial
plane in the direction along the dipole magnetic line to the star surface.
And the accretion flow will be modulated by the frequency $\omega
_{j}(f_{1})+\Omega $. The point $f=f_{1}$ will be factually the inner edge
of the accretion disk. In observed X-ray emission we can see several
characteristic frequencies:
\begin{enumerate}
\item $\omega _{\mathrm{K}}(f_{1})$, it is the modulation of the X-ray flux by the
inhomogeneity of the inner edge of the disk;

\item $\omega _{\mathrm{K}}(f_{1})-\Omega $, it is the beating frequency corresponding
to the scattering of X-rays from the sources on the star surface, rotating
with $\Omega $ , by the inner edge of the disk;

\item $\omega_j(f_1)+\Omega =i\cdot\omega _{\mathrm{K}}(f_{1})+(1-i)\cdot \Omega $,
 it is the frequency of the accretion flow modulated by Alfven oscillations;

\item different combinations of frequencies 1, 2, 3.

\end{enumerate}

\section{Sonic oscillation}

\begin{figure}
\includegraphics[width=88mm]{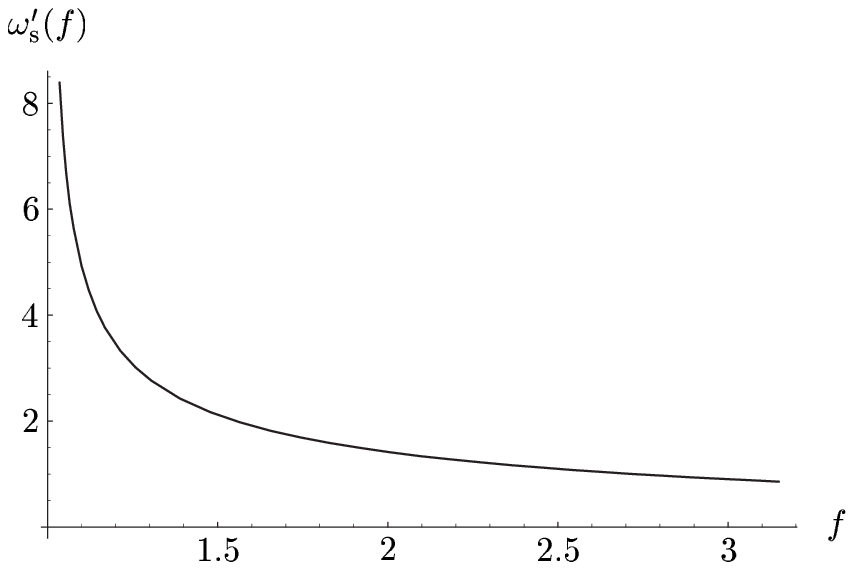}
\caption{The dependence of sonic first harmonics $\omega'_{\mathrm s}$ on $f$.}
\label{sonic}
\end{figure}

It is naturally to consider that the plasma flow along the
tube $f=f_1$ is not stationary. The sound mode can be excited.

Let us consider plasma as isothermic with temperature $T$.
For disturbance of concentration $\delta n$
along the magnetic tube we receive from hydrodynamics
\begin{equation}
\delta n+\frac T{m_{\mathrm p}\omega^2}\Div\grad\delta n=0.
\end{equation}
In dipole coordinates it looks like
\begin{equation}
\delta n+\frac T{m_{\mathrm p} \omega^2}B^2\frac{\partial^2}{\partial l^2}\delta n=0.
\end{equation}
The boundary condition is on the star surface. The plasma velocity
component which is normal to the surface must be equal to zero.
\begin{equation}
\vec v=-\left.i\frac{T}{m_{\mathrm p}n_{\mathrm p}\omega}\vnabla\delta n\right|_{l=\pm l_m}=0.
\end{equation}

The first harmonics corresponding to the solution of
this equation with boundary condition can be presented as
\begin{equation}
\omega(f)=\omega_{\mathrm s}^{\strut}\cdot\omega'_{\mathrm s}(f),
\end{equation}
where $\omega_{\mathrm s}^2=T/m_{\mathrm p}R^2$ and $\omega'_{\mathrm s}(f)$
is dimensionless
function on $f$, which is presented on the Fig.~\ref{sonic},
$\omega'_{\mathrm s}(f)$ is of the order of $1$ when $f\sim2$.
In the region of $f\approx1$ the function $\omega'_{\mathrm s}(f)$ can be approximated
by the expression $2/(\pi\sqrt{f-1})$.
So, $\omega_0$ is the charachteristic frequency for
sonic oscillation.
If we take $T=1$~keV and $R=10$~km then $\omega_0=30$~s$^{-1}$
or $\nu_0=4.9$~Hz. Thus, sonic oscillation can be found on
$10$~Hz band that is observed simaltaneously with the QPO
of $1$~kHz band.

\section{Lense-Thirring precession}
\begin{figure}
\includegraphics[width=88mm]{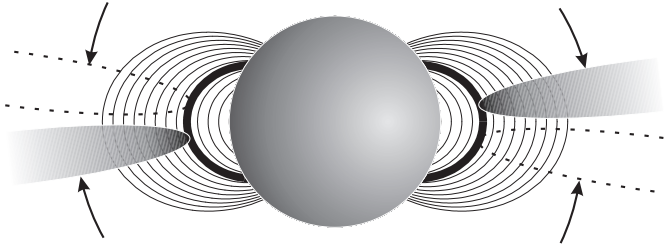}
\caption{Due to Kerr metrics the disk precesses with Lense-Thirring frequency.}
\label{lense}
\end{figure}

While the accreting disk is spinning at the distance $r$ about the neutron star
with Keplerian frequency, it precesses about its
equilibrium position
with Lense-Thirring frequency due to Kerr metrics \citep{thorne, morsink}
\begin{equation}
\omega_{\mathrm {LT}}=a\frac{r_{\mathrm g}}{r^3+\left(\displaystyle a/c\right)^2(r_{\mathrm g}+r)},
\end{equation}
where $a=J/M$ ($J=I\Omega$ is the star's angular momentum and
$I=2MR^2/7$ is the momentum if inertia of the star),
$r_{\mathrm g}=2GM/c^2$ is the gravitational radius of the neutron star of mass $M$
and radius $R$.

In the system of units where $M$ is measured in masses of sun
and $R$ in $10$~km the Lense-Thirring frequency
can be presented in the form of
\begin{equation}
\omega_{\mathrm {LT}}=8.44\cdot10^{-2}~\frac{M\Omega}{R}f^{-3}.
\end{equation}

For $M/R\simeq1$, $f\simeq1$ and $\Omega\simeq2\pi\cdot300$~s$^{-1}$
$\omega$ is of the order of $2\pi\cdot25$~s$^{-1}$.
The second harmonics of Lense-Thirring precession is usually observed
\citep{stella}. In our model it can be illustrated by the
fact that some mass is thrown out the disk each time it is
in extreme position, i.e. with frequency
$2\omega_{\mathrm {LT}}$ (see Fig.~\ref{lense}).

\section{Scorpius X-1}
We have applied our speculations for Sco X-1 \citep{klis2}.
There were observed 4 QPOs:
$6$--$20$~Hz normal/flaring quasi-periodic oscillations;
$45$~Hz QPOs and two kHz QPOs.
According to our model we ascribed the $6$--$20$~Hz QPOs to
the sound oscillations, $45$~Hz QPOs to Lense-Thirring oscillations,
and two kHz QPOs to Alfven and Keplerian oscillations correspondingly.

Below we use the system of units where temperature is measured in eV,
masses of objects~-- in masses of sun and distances~-- in $10$~km.
Thus, we can write
\begin{align}
\text{for sonic:}&&
        \omega_{\mathrm s}(f)&=0.98~\sqrt{\frac T{R^2}}~\omega'_{\mathrm s}(f),\\
\text{for Keplerian:}&&
        \omega_{\mathrm{K}}(f)&=11.5\cdot10^3~\sqrt{\frac M{R^3}}~f^{-3/2}_{\strut},
\end{align}
where $f=r/R$~-- a parameter that characterizes the distance
between the disk and neutron star surface.

\begin{figure}
\includegraphics[width=88mm]{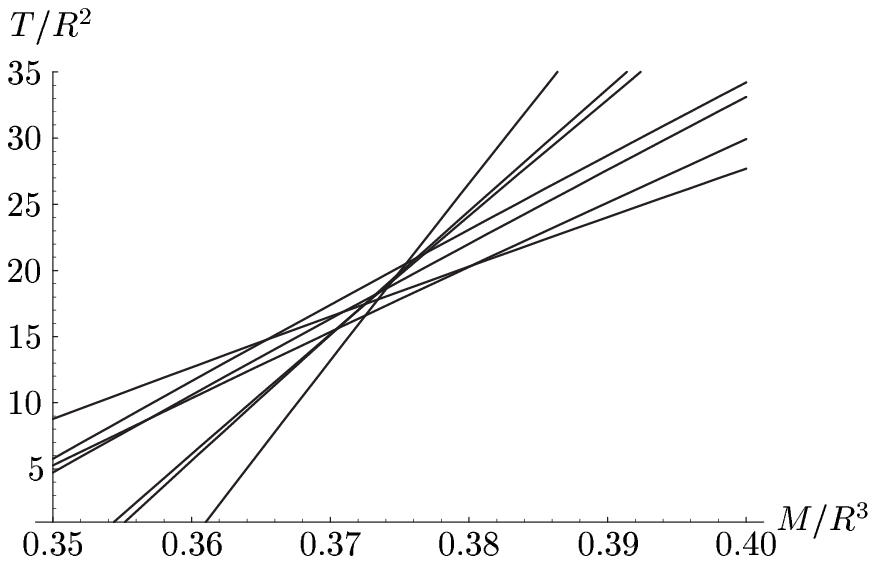}
\caption{Each line on this graph corresponds to one measurment
of sonic and Keplerian frequencies. $T$ is measured in eV, $M$ in $\Msun$,
$R$ in $10$~km.}
\label{scox1}
\end{figure}

For every pair of data $(\omega_{\mathrm{K}},\omega_{\mathrm s})$ we can build a
curve in $\left(\dfrac M{R^3};{\dfrac T{R^2}}\right)$ space.
This results in the graph on Fig.~\ref{scox1}.

The axis of abscissae is the space for possible meanings of
the ratio of $M/R^3$. For each of possible meanings of this ratio
we can find the temperature $T$ (or, to be more accurate,
the ratio $T/R^2$) for each experiment on determining
the frequencies $\omega_{\mathrm{K}}$ and $\omega_{\mathrm s}$.
So we can evaluate the range of the change of temperature $T$.
One can find it remarkable that all the curves cross
each other in almost one point!
So we can suppose that the temperature of the disk is approximately
constant.
From the graph we determine $M/R^3=0.374\pm0.004$ and $T/R^2=16.8\pm0.8$.

\begin{figure}
\includegraphics[width=88mm]{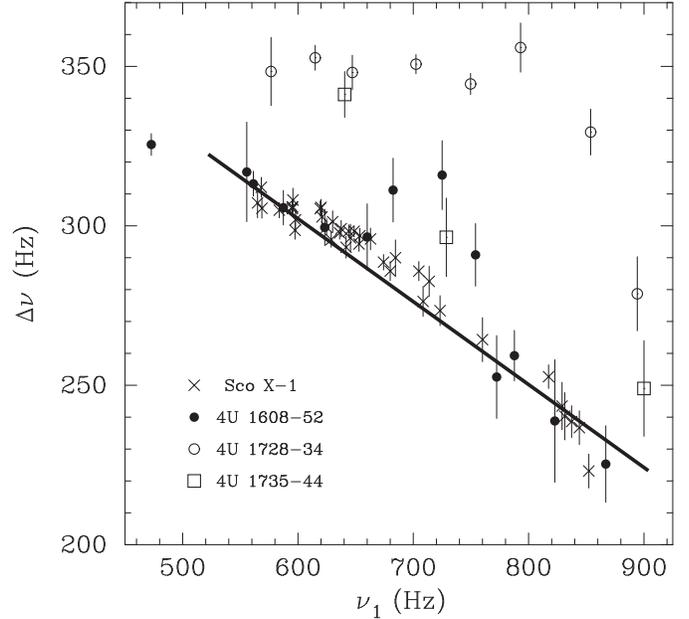}
\caption{The variations in kHz QPO peak separation as
a function of the lower kHz frequency \citep{mendez}.
It is well-seen that for Sco X-1 the dependence is linear.}
\label{menline}
\end{figure}

\cite{mendez} have observed that the separation
between the two high-frequency QPO peaks changes
from one observation to another.
In our work we considered the equation
\begin{equation}
i\cdot (\omega_{\mathrm{K}}-\Omega)=\omega_j\simeq j\cdot\omega_{\mathrm A}.
\end{equation}
In general case the peak separation
will be
\begin{equation}
\omega_{\mathrm{K}}-\omega_{\mathrm A}=
\frac ji\omega_{\mathrm{A}}+\Omega-\omega_{\mathrm{A}}=
\left(\frac ji-1\right)\omega_{\mathrm{A}}+ \Omega.
\label{eq30}
\end{equation}
It is easy to see, that when $i\neq j$ the peak
separation will not remain constant.
For example, if $i=2$ and $j=1$ then
$\omega_{\mathrm{K}}-\omega_{\mathrm A}=-\omega_{\mathrm{A}}/2+\Omega$.
If $i=1$ and $j=2$, then
$\omega_{\mathrm{K}}-\omega_{\mathrm A}=\omega_{\mathrm{A}}+\Omega$.
Here we must mention that Alfven eigen frequency
depends on the parameter $\lambda$ that changes when
$f$ will change and, therefore, when $f$ changes,
numbers $i$ and $j$ can change.
As a result, the dependence of the peak separation on $\omega_{\mathrm{K}}$
will be a piece linear function.
The dependencies of separation frequency on the lower one
for different objects is
shown on the Fig.~\ref{menline}.
For Sco X-1 the dependence can be well
approximated by the linear function \eqref{eq30} with
$j/i=3/4$ and $\Omega=2\pi\cdot450$~s$^{-1}$.

\begin{figure}
\includegraphics[width=88mm]{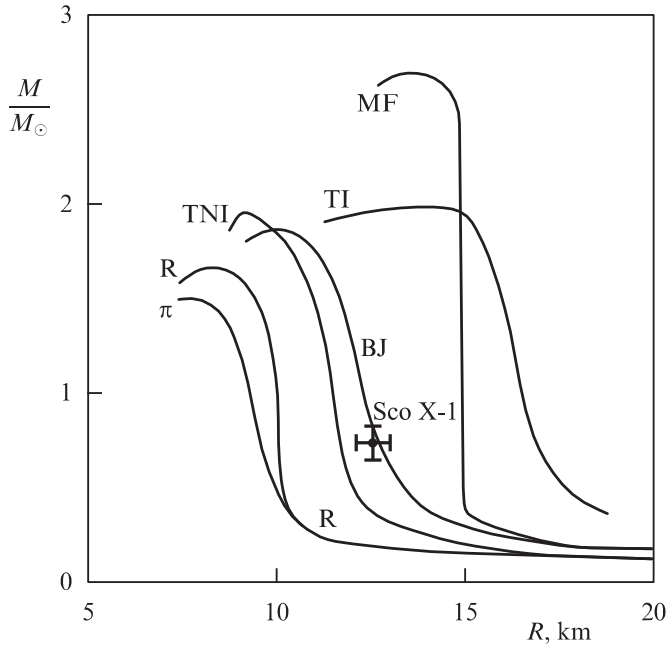}
\caption{
Diagram of the dependence of gravitational mass on the radius
for different equations of state of nuclear matter \citep{baym}
and our point for Sco X-1 on it. Larger neutron star
radii are given by stiff (TI, MF) rather than soft (R, $\pi$) equations of state.
The classification corresponds to \cite{alcock}.
}
\label{point}
\end{figure}


So, comparing the $45$~Hz QPO with Lense-Thirring precession we determine
the ratio $M/R\approx0.60.$
Finally $M=0.75\Msun$, $R=12.6$~km, $T=27$~eV.
The point $(M,R)$ on the gravitational mass -- radius diagram 
is shown on the Fig.~\ref{point}.

For this $M,R,T$ we can find that
$\omega_{0\mathrm s}=4.16$~s$^{-1}$ or $\nu_{0\mathrm s}=0.66$~Hz.
This low value can be explained by the fact
that disk approaches too close to the neutron star
(down to $f=1.01$) and in this region of $f$
function $\omega_{\mathrm s}'(f)$ takes values much greater than $1$
(e.g. $\omega'_{\mathrm s}(1.01)=15.7$).

\begin{figure}
\includegraphics[width=88mm]{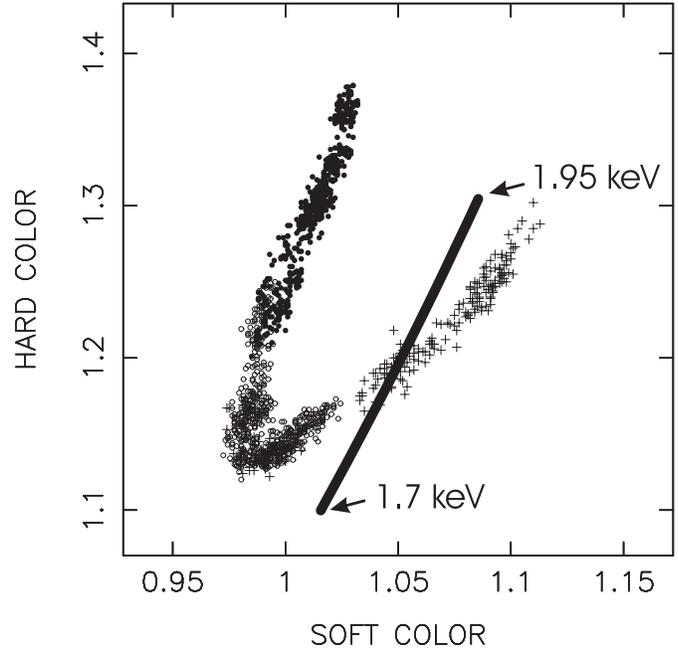}
\caption{
The X-ray color-color diagram for Sco X-1 \citep{klis2}.
Soft color is the (3-5~keV)/(1-3~keV),
hard color the (7-20~keV)/(5-7~keV), count rate ratio.
The bold line corresponds to
the black body radiation with temperature varying from $1.7$~keV to $1.95$~keV.}
\label{softhard}
\end{figure}

At first sight, $T$ must be equal to the radiation temperature
of the neutron star, which can be determined from the
X-ray color-color diagram. We put it on one graph from
Scorpius X-1 and from black body (see Fig.~\ref{softhard}).
The normal branch of that from
Scorpius almost coincide with that of black body.
And the temperature of the black body varies from $1.7$~keV up to $1.95$~keV.
But if we estimate the free path of the photons from the neutron star
$\langle l\rangle\sim1/\sigma n_{\mathrm p}\simeq20$~km we find that it is
greater then $R$ and so, radiation does not heat
the inner part of the disk.

\begin{acknowledgements}
This work is supported by Russian Fundamental Research Foundation
Grant No 99-02-17184.
\end{acknowledgements}

\bibliographystyle{apj}

\begin{thebibliography}{11}
\expandafter\ifx\csname natexlab\endcsname\relax\def\natexlab#1{#1}\fi

\bibitem[{Alcock {et~al.}(1986)Alcock, Farhi, \& Olinto}]{alcock}
Alcock, C., Farhi, E., \& Olinto, A. 1986, ApJ, Vol. 310, 261

\bibitem[{Baym \& Pethick(1979)}]{baym}
Baym, G. \& Pethick, C. 1979, ARA\&A, Vol.~17, 415

\bibitem[{Klein {et~al.}(1998)Klein, Jernigan, Arons, Morgan, \& Zhang}]{klein}
Klein, R.~I., Jernigan, J.~G., Arons, J., Morgan, E.~H., \& Zhang, W. 1998,
  ApJ, Vol. 508, 791--830

\bibitem[{M\'endez \& van~der Klis(1999)}]{mendez}
M\'endez, M. \& van~der Klis, M. 1999, ApJ, Vol. 517, L51--54

\bibitem[{Miller {et~al.}(1998)Miller, Lamb, \& Psaltias}]{miller}
Miller, M.~C., Lamb, F.~K., \& Psaltias, D. 1998, ApJ, Vol. 508, 791--830

\bibitem[{Morsink \& Stella(1999)}]{morsink}
Morsink, S.~M. \& Stella, L. 1999, ApJ, Vol. 513, 827--44

\bibitem[{Stella \& Vietri(1999)}]{stella}
Stella, L. \& Vietri, M. 1999, Phys. Rev. Lett., Vol.~82, 17

\bibitem[{Thorne {et~al.}(1986)Thorne, Price, \& Macdonald}]{thorne}
Thorne, K.~S., Price, R.~H., \& Macdonald, D.~A. 1986, Black Holes: The
  Membrane Paradigm (Yale University Press)

\bibitem[{Titarchuk \& Osherovich(1999)}]{titarchuk}
Titarchuk, L. \& Osherovich, V. 1999, ApJ, Vol. 518, L95--98

\bibitem[{van~der Klis(2000)}]{klis1}
van~der Klis, M. 2000, ARA\&A, Vol.~38, 717--760

\bibitem[{van~der Klis {et~al.}(1996)van~der Klis, Swank, Zhang, Jahoda, \&
  Morgan}]{klis2}
van~der Klis, M., Swank, J.~H., Zhang, W., Jahoda, K., \& Morgan, E.~H. 1996,
  ApJ, Vol. 469, L1--4

\end{thebibliography}

\end{document}